\documentclass[prb,twocolumn,superscriptaddress,showpacs,floatfix]{revtex4}

\usepackage{epsfig,amsmath,amssymb,latexsym}

\newcommand{\beq}{\begin{eqnarray}}
\newcommand{\eeq}{\end{eqnarray}}
\newcommand{\siesta}{\textsc{Siesta}}

\usepackage{subfigure}

\begin{document}
\newcommand{\Go}{\ensuremath{\textrm G_{0}}}
\newcommand{\brho}{\mbox{\boldmath$\rho$}}

\title{DFT study of graphene antidot lattices: The roles of geometry relaxation and spin}
\preprint{1}
\author{Joachim A. F\"{u}rst}
\email[Corresponding author: ]{joachim.fuerst@nanotech.dtu.dk} 
\affiliation{DTU
Nanotech -- Department of Micro- and Nanotechnology, Technical
University of Denmark, DK-2800 Kongens Lyngby, Denmark}
\author{Thomas G. Pedersen}
\affiliation{Department of Physics and Nanotechnology, Aalborg University, DK-9220 Aalborg East, Denmark}
\author{Mads Brandbyge}
\affiliation{DTU Nanotech -- Department of Micro- and
Nanotechnology, Technical University of Denmark, DK-2800 Kongens
Lyngby, Denmark}
\author{Antti-Pekka Jauho}
\affiliation{DTU Nanotech -- Department of Micro- and
Nanotechnology, Technical University of Denmark, DK-2800 Kongens
Lyngby, Denmark} \affiliation{Department of Applied Physics,
Helsinki University of Technology, P. O. Box 1100, FI-02015 HUT,
Finland}
\date{\today}
\begin{abstract}
Graphene sheets with regular perforations, dubbed as antidot
lattices, have theoretically been predicted to have a number of
interesting properties. Their recent experimental realization with
lattice constants below 100 nanometers stresses the urgency of a
thorough understanding of their electronic properties. In this work
we perform calculations of the band structure for various
hydrogen-passivated hole geometries using both spin-polarized
density functional theory (DFT) and DFT based tight-binding (DFTB)
and address the importance of relaxation of the structures using
either method or a combination thereof. We find from DFT that all
structures investigated have band gaps ranging from 0.2 eV to 1.5
eV. Band gap sizes and general trends are well captured by DFTB with
band gaps agreeing within about 0.2 eV even for very small structures. A
combination of the two methods is found to offer a good trade-off
between computational cost and accuracy. Both methods predict
non-degenerate midgap states for certain antidot hole symmetries.
The inclusion of spin results in a spin-splitting of these states as
well as magnetic moments obeying the Lieb theorem. The local spin
texture of both magnetic and non-magnetic symmetries is addressed.
\end{abstract}

\maketitle
\section{Introduction}
Graphene, the single-atom thick two-dimensional sheet of carbon
atoms, has stimulated considerable experimental \cite{Geim2007} and
theoretical research \cite{Katsnelson2006} as well as proposals for
future nanodevices\cite{Avouris2007}. Various graphene-based
applications have been realized in recent
years\cite{Novoselov2005,Novoselov2007} and the relevance for
application in devices is heavily increased by the rapidly improving
ability to pattern monolayer films with e-beam lithography
\cite{Geim2007} where features on the ten-nm scale have been
obtained \cite{Han2007,Fischbein2008}. Moreover, recent advances in 
chemical vapor deposition of graphene (see e.g. Ref. [\onlinecite{Kim2009}])
are promising for fabrication of large area, high quality devices.

Yet another way of nanoengineering graphene consists of defining an antidot lattice on graphene
by means of a regular array of nanoscale perforations. This theoretical idea was introduced by
Pedersen {\it et al.} \cite{Pedersen2008,Pedersen2008b}, who showed using
tight-binding calculations that antidot lattices change the
electronic properties from semimetallic to semiconducting with a
significant and controllable band gap. Such structures have recently
been realized experimentally by Shen {\it et al.} \cite{Shen2008} and Eroms {\it et al.}
\cite{arxiv.0901.0840v1} with lattice spacings down to 80 nm.
Quantum dots and graphene ribbons have been demonstrated with
dimensions of only a few nm \cite{Ritter2009}. Very recently, Girit
{\it et al.} \cite{Girit2009} have studied the dynamics at the edges
of a growing hole in real time using a transmission electron
microscope. Both in the experiment and in Monte Carlo simulations
they find the zig-zag edge formation to be the most stable
structure. This is in agreement with the findings of Jia {\it et
al.}\cite{Jia2009} who
demonstrate a method to produce graphitic nanoribbon edges in a
controlled manner via Joule heating. This opens the possibility of
making antidot lattices with a desired hole geometry.
\begin{figure}[tbh]
\begin{center}
\mbox{
\subfigure{}\includegraphics[width=0.48 \columnwidth,angle=0,viewport=160 55 410 280,clip]{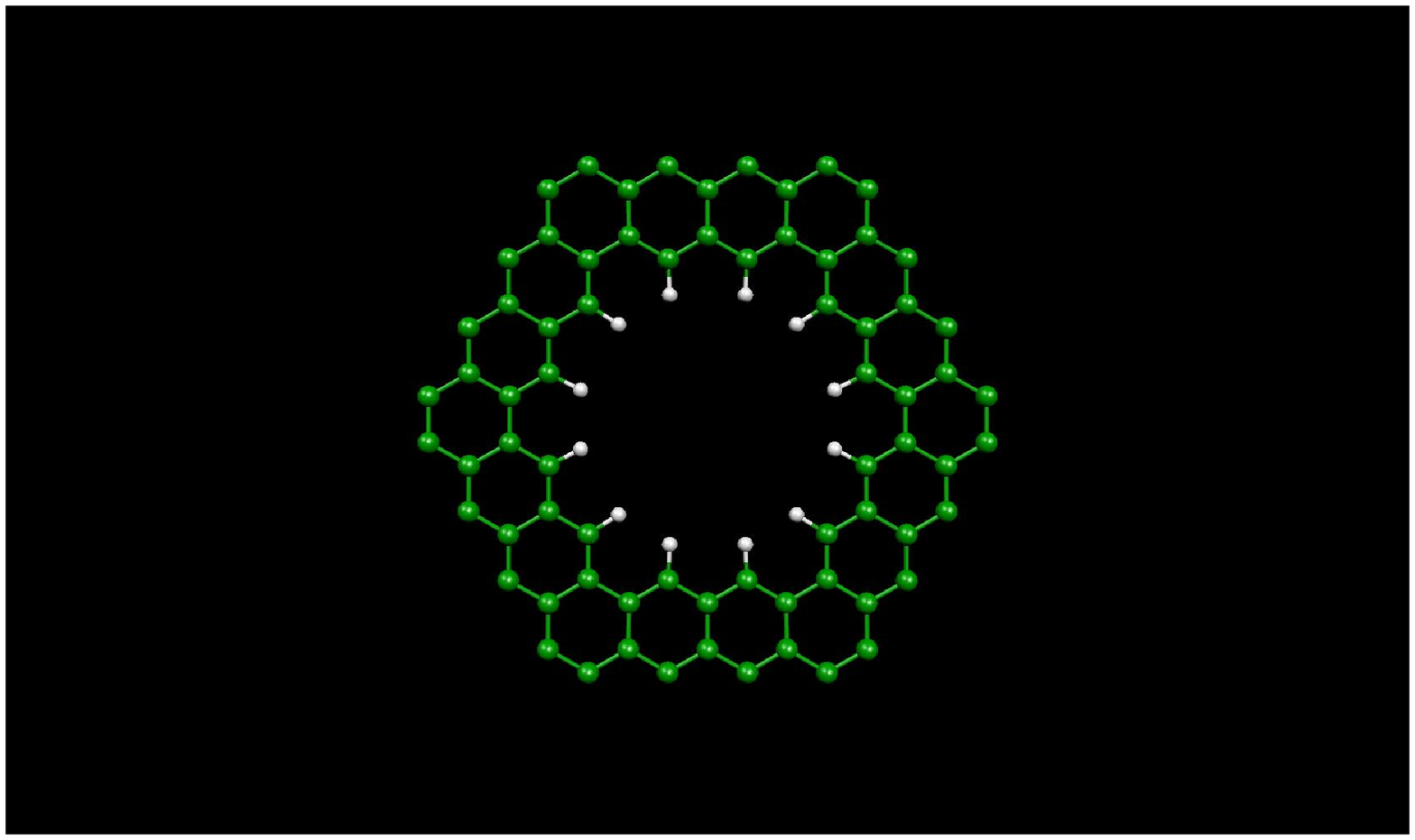}
\subfigure{}\includegraphics[width=0.474 \columnwidth,angle=0,viewport=100 0 470 355,clip]{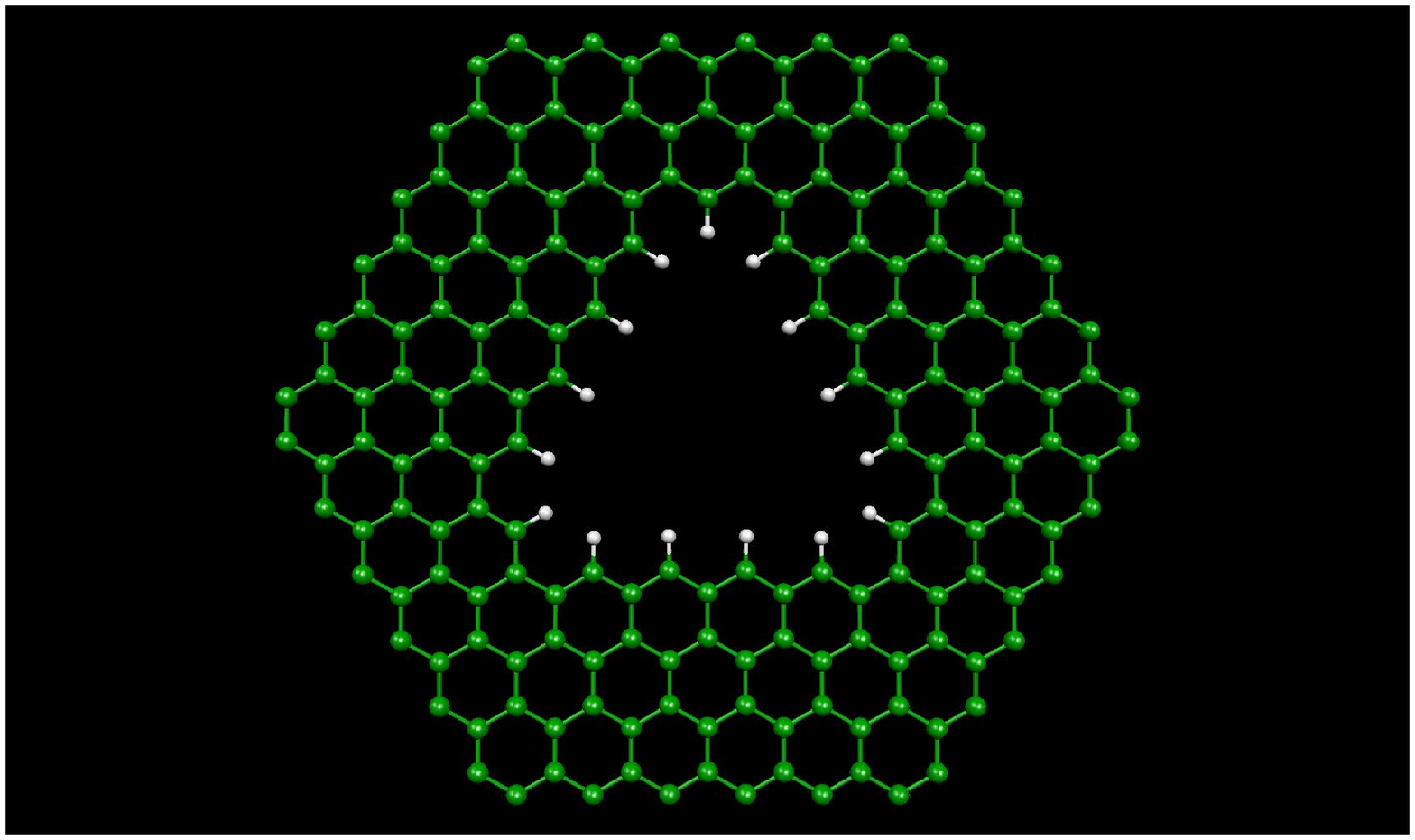}
}
\end{center}
\caption{ (Color online) The unit cell of the $\{4,2\}_\bigcirc$(left) and $\{6,5\}_\triangle$
(right) system. The hexagonally shaped unit cells are repeated in plane to
form a honeycomb lattice of antidots. The carbon atoms (green) are hydrogen
terminated (white) along the hole edges. } \label{unitcells}
\end{figure}

While carbon nanotubes, graphene and recently graphene ribbons have
been studied extensively using first principles methods, antidot
lattices  in graphene have mainly been treated with simpler
models\cite{Pedersen2008,arxiv.0903.0918v1}. The very recent work by Vanevic {\it et al.} \cite{arxiv.0903.0918v1} uses a $\pi$-orbital tight-binding model to study antidot lattices with rather large lattice constants (these systems are more easily accessed experimentally but cannot be analyzed in terms of {\it ab initio} methods). Their focus is on the possible occurrence of midgap states without introducing defects in the antidot lattice, as was the case in the original proposal by Pedersen {\it et al.}
 
Many studies on magnetization have been reported for various
graphene
structures\cite{Lieb1989,Lehtinen2004,Palacios2008,Kumazaki2008,Jiang2007,Philpott2008}.
The origin of the magnetism can be understood based on the theory by
Lieb \cite{Lieb1989}, and the subsequent related work by Inui {\it
et al.} \cite{Inui1994} on the properties of the bipartite lattice.
Single vacancies and their spin properties have been studied by,
e.g. Lehtinen {\it et al.} \cite{Lehtinen2004} and Palacios {\it et
al.}\cite{Palacios2008}; the latter paper also investigates voids in
both graphene and graphene ribbons in detail using a mean-field
Hubbard-model. Magnetization has also been studied in carved
slits\cite{Kumazaki2008}, finite ribbons\cite{Jiang2007} and flakes
\cite{Jiang2007,Philpott2008} as well as rings \cite{Bahamon2009}
and notches \cite{Hancock2008}. Recently, DFT treatments of magnetic
properties of nano-holes in graphene\cite{Yu2008} and graphene films
\cite{Chen2008} have been published.

The realized antidot lattices with hole sizes of several tens of nanometers
and even larger lattice spacings involve several thousands of atoms in a unit cell 
and are computationally too costly to be treated with DFT in a
systematic manner. The DFT based tight-binding method, DFTB
\cite{Porezag1995}, however, allows one to address such large
systems. The difference in computational cost between DFT and DFTB
is for the present study found to be at least a factor of thirty. We
thus investigate the accuracy of DFTB compared to DFT on
much smaller antidot lattices in terms of the band
structures\footnote{We note, that DFT is known to underestimate the
band gap, see, e.g. M.S.Hybertsen and S.G. Louie, Phys. Rev. B
\textbf{34}, 5390 (1986).}. Since geometry relaxation is the most
costly task in DFT we also investigate the cost benefits of
combining the two methods. By using DFT and elaborating on the role
of spin, we also wish to address some of the main features found
specifically for antidot lattices on a tight-binding level of
the theory.

This paper is organized as follows. In Sec. \ref{sec_method} we
introduce the antidot lattice systems and the methods used. The
equilibrium geometries and band structures obtained using both DFT
and DFTB and a combination thereof is given in Sec. \ref{sec_res}
together with a detailed investigation of the spin properties. We conclude with a short summary.
\section{Systems and Method}\label{sec_method}

The specific realization of the antidot lattice we consider in this
paper is a hexagonal (triangular) array of holes in a graphene sheet
as proposed by Pedersen $\textit{et al.}$ \cite{Pedersen2008}.
Within the hexagonal unit cells there can be different hole
geometries, and two examples of high symmetry holes are shown in
Fig. \ref{unitcells}. Below, geometries are fully relaxed but as a
starting point ideal geometries using fixed bond lengths and angles 
of $120^\circ$ are constructed. These geometries furthermore provide
a straightforward notation for the structures. Thus, we designate antidot 
lattices with circular holes according to the notation
$\{L,R\}_\bigcirc$, where $L$ is the side length of the unit cell and $R$ 
the hole radius, both measured in units of the graphene lattice 
constant $a$=2.46 \AA ~giving a C-C bond length of 1.42 \AA. 
Similarly, for triangular holes, we apply the notation $\{L,D\}_\triangle$, 
where $D$ denotes the side length of the hole \cite{arxiv.0903.0918v1}. The holes are passivated 
with H using a C-H bond length of 1.1 \AA ~and consist almost entirely 
of zig-zag edges. These structures are idealized but may well be within 
experimental reach given the recent advancements \cite{Girit2009,Jia2009}.

For the first principles calculations we have used the {\it ab
initio}  pseudopotential DFT as implemented in the \siesta
~code\cite{siesta} to obtain the electronic structure and relaxed
atomic positions from spin-polarized DFT\footnote{The systems are
relaxed using the conjugate gradient (CG) method with a force
tolerance of 0.01 eV/{\AA}. All atoms are allowed to move during
relaxation but the unit cell is kept fixed with a vacuum between
graphene layers in neighboring cells of 10 \AA. The mesh cutoff
value defining the real space grid used is 175 Ry and we employ a
double-$\zeta$ polarized (DZP) basis set. A Monkhorst-Pack grid of
(2,2,1) was found sufficient for all structures. }. We employ the
GGA PBE functional for exchange-correlation\cite{Perdew1996}.

For the DFTB results\footnote{We relax the two outermost  rows of
atoms along the perimeter of the hole until the total energy per
atom was converged to better than $10^{-4}$ eV.}, we use the
original (C,H) parametrization of Porezag {\it et
al.}\cite{Porezag1995} which does not include spin.

\section{Results}\label{sec_res}
\begin{table}
    \centering
        \begin{tabular}{| l|c|c |c|c|c |c|}
            \hline \hline
             &\multicolumn{3}{|c|}{DFT}&\multicolumn{1}{|c|}{Non-spin DFT}&\multicolumn{2}{|c|}{DFTB}\\
            \hline
           Relaxation& None & DFT& DFTB & DFTB & None & DFTB \\  \hline
           $\{4,2\}_\bigcirc$ &0.93 & 1.01 & 0.97 &0.97 &0.99&1.05 \\ \hline
       $\{5,2.8\}_\bigcirc$ &0.72 &0.84 & 0.79 & 0.79&0.88&0.98 \\ \hline
       $\{5,3.5\}_\bigcirc$ &1.27 &1.51 & 1.35 & 1.35&1.72&1.74 \\ \hline
       $\{6,3.6\}_\bigcirc$ &0.39 &0.52 & 0.46& 0.46&0.65&0.75 \\ \hline
       $\{6,5\}_\triangle$ &0.24 &0.22 & 0.22 & 0.00&0.00&0.00 \\ \hline \hline
        \end{tabular}
    \caption{The band gaps for various systems calculated with either
    DFT or DFTB using geometries obtained with different methods for relaxation.
    All values are in eV.}
    \label{bandgaps}
\end{table}
For all considered structures a structural relaxation with DFT leads
to a shrinking of the hole, of the order of 1 \%, resulting in C-C
bonds close to the edges stretching and contracting in the range of
1.39 - 1.45 (1.43 for $\{5,3.5\}_\bigcirc$) \AA. A few bond lengths away from
the hole edge the C-C bond length remains unaltered at 1.42 \AA.

In the case of relaxation with DFTB the picture is quite  similar.
Edge-atom C-C bond lengths vary from 1.39-1.42 \AA~for all systems
but $\{6,5\}_\triangle$, which has variations of 1.38 - 1.44 \AA. The
shrinking of the hole size is smaller than 1\%.
\begin{figure}[tbh]
\begin{center}
\includegraphics[width=0.75 \columnwidth,angle=-90,viewport=70 30 570 740,clip]{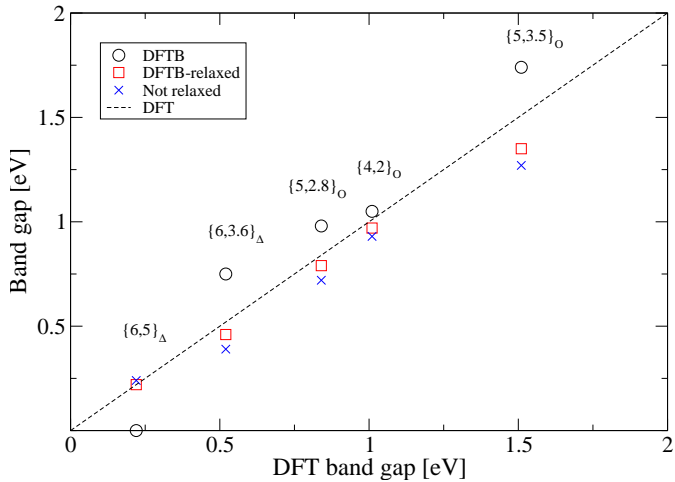}
\end{center}
\caption{(Color online) The band gaps for DFTB, DFT on DFTB-relaxed
geometry and  DFT on unrelaxed geometry plotted versus  pure DFT
results. Points above (below) the dotted line are thus overestimated
(underestimated) compared to pure DFT. Note, that DFTB calculates the electronic structure without spin and fails to predict a band gap for the $\{6,5\}_\triangle$ structure.} \label{methodbandgaps}
\end{figure}

The results for the band gaps for five different systems are
summarized  in Table \ref{bandgaps}. The band structures are
calculated using both DFT and DFTB on structures relaxed at
different accuracy levels and thus at different computational costs.
The combinations are not exhaustive but represent a relevant set
aimed at saving computational costs. The relaxation type is given in
the second row of Table \ref{bandgaps}. DFTB is expected to match
DFT better as $L/R$ increases due to decreased importance of edge
details. The systems chosen here are mostly edge-dominated (low
$L/R$ ratio) and thus represent the worst case scenario.

Using DFT we find band gaps ranging from 0.2 to 1.5 eV confirming
that the antidot lattice turns the semimetallic graphene into a
semiconductor \cite{Pedersen2008}. However, only spin-polarized DFT
predicts a band gap for the $\{6,5\}_\triangle$ structure which will be
discussed in detail below.

Pedersen \textit{et al.} \cite{Pedersen2008} demonstrated a scaling-law
between the hole size and the band gap for large $L/R$ ratios but no
such simple picture for small $L/R$ ratios emerged. This trend
agrees well with our edge-dominated systems were no simple scaling
between the hole size and the band gap could be identified.

As illustrated on Fig. \ref{methodbandgaps}, DFTB in general gives a
larger band gap than DFT (the circles lie consistently above the
dashed line). This tendency is enhanced the more edge-like the
structure becomes. On average, the DFTB gap is 20\% larger than the DFT value
for the four structures with circular perforations. The discrepancy increases 
for structures with holes occupying a large portion of the unit cell, such as 
$\{5,3.5\}_\bigcirc$. Moreover, for $\{6,5\}_\triangle$ the omission of
spin effects leads incorrectly to a vanishing DFTB band gap in agreement with 
non-spin DFT. In
the cases of the $\{4,2\}_\bigcirc$ and $\{6,5\}_\triangle$ systems the band
structures are shown in Fig. \ref{bands} left and right,
respectively, calculated using DFT (DFTB) in the upper(lower) panel.
We see that the shape of the bands corresponds qualitatively for the
two methods.
\subsection{Relaxation}
We next analyze the importance of relaxation. The clear trend is
that relaxation increases the band gap. This is illustrated for DFT
in Fig. \ref{methodbandgaps} (crosses corresponding to the unrelaxed structure lie below the dotted line), as well as in Table \ref{bandgaps} for DFTB. Comparing
unrelaxed results with fully relaxed results we see from Table
\ref{bandgaps} a change in band gaps within 10 \% and 15 \% using
DFTB and DFT, respectively. Only in the case of DFT does the effect
of relaxation increase the more edge-dominated the system becomes.
The change using DFT for the $\{4,2\}_\bigcirc$ system is 8 \% compared to
16 \% for $\{5,3.5\}_\bigcirc$.

It must be emphasized that larger differences between initial and
relaxed geometries may well give rise to a larger discrepancy between their band gaps. However, even for the case of a single
passivated edge-defect the difference in relaxed and unrelaxed DFT
band gaps is less than 10 \%.
\begin{figure}[tbh]
\begin{center}
\includegraphics[width=0.8 \columnwidth,angle=-90,viewport=20 50 550 750,clip]{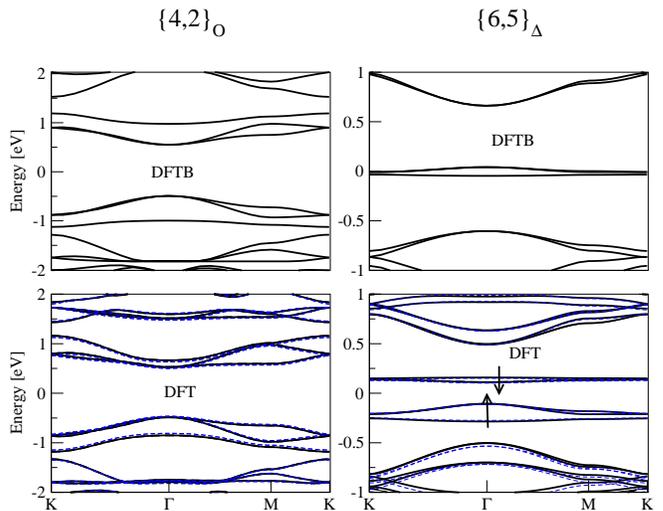}
\end{center}
\caption{(Color online) Band structures for the $\{4,2\}_\bigcirc$  (left
column) and $\{6,5\}_\triangle$ (right column) systems using DFTB (upper
panels) and DFT (lower panels). The dotted blue curves in the lower
panels are DFT results using structures relaxed with DFTB. The arrow
pointing up (down) indicates that bands are filled with majority
(minority) electrons only.} \label{bands}
\end{figure}

By relaxing the geometry with DFTB the DFT results are  improved
from the non-relaxed case as shown in Fig. 2 (the squares are closer
to the dashed line than the crosses). Compared to pure DFT the
largest difference in the band gap is again found for the
edge-dominated $\{5,3.5\}_\bigcirc$ system: it is now 11 \% compared to 16 \%
without relaxation. For the larger $\{6,5\}_\triangle$ structure we find the
same values as for pure DFT. The DFT results are shown for both
DFT-relaxed and DFTB-relaxed structures on Fig. \ref{bands}
indicated by thick and dotted line, respectively. The different
geometries do not change the bands notably.

Using DFT on DFTB-relaxed structures is thus an approach with a good
trade-off between accuracy and computational cost. This finding is
of great practical use, since relaxation is very costly in DFT.

\subsection{Magnetic properties}

For the $\{6,5\}_\triangle$ system with both non-spin-DFT and DFTB there are
three (one nearly doubly degenerate) bands with weak dispersion at
zero energy. Introducing spin leads to a clear splitting of these
bands, i.e., to the formation of  a band gap.  For a comparison of
DFT and DFTB, see Fig. \ref{bands}, right column. The three bands
below (above) the Fermi level are half-filled by majority (minority)
spin electrons and are thus completely spin-polarized. The size of
the band gap is thus also an indication of the robustness of the
magnetic
state \cite{Kusakabe2003}.\\
\begin{figure}[tbh]
\begin{center}
\includegraphics[width=0.95 \columnwidth,angle=0,viewport= 90 10 640 520, clip]{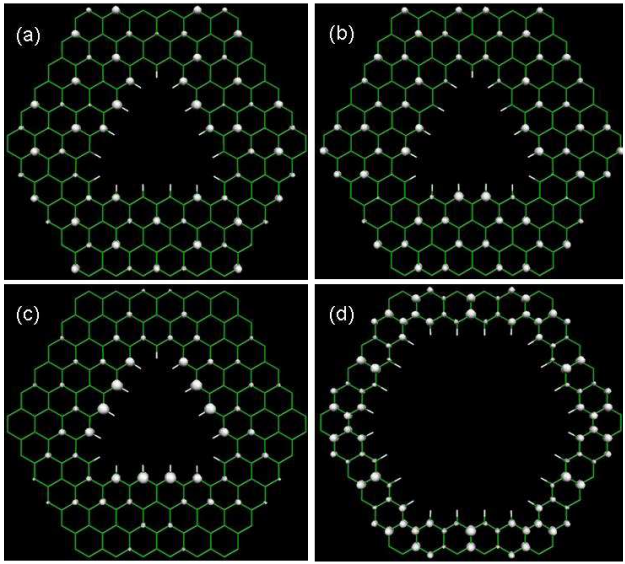}
\end{center}
\caption{(Color online) Amplitudes of certain important states.
(a,b): degenerate midgap states, (c): non-degenerate midgap state of
$\{6,5\}_\triangle$, (d): highest filled bands of $\{6,3.6\}_\bigcirc$. All states
are calculated at the $\Gamma$-point.} \label{prob}
\end{figure}
\begin{figure}[tbh]
\begin{center}
\includegraphics[width=0.95 \columnwidth,angle=0,viewport= 165 70 400 350, clip]{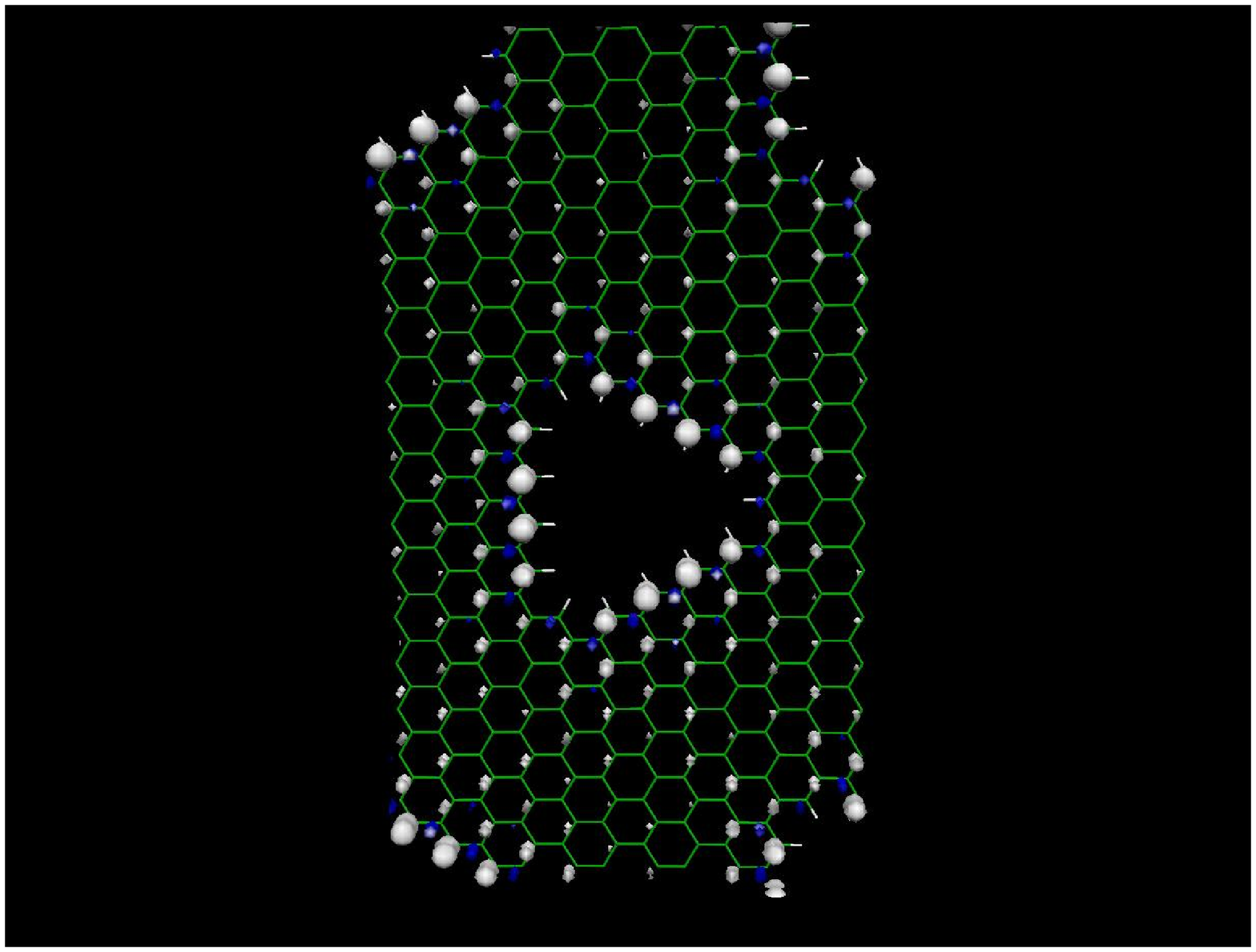}
\end{center}
\caption{(Color online) The difference in majority spin and minority
spin densities for the $\{6,5\}_\triangle$ structure. Blue (white) indicates
surplus of majority(minority) spin.} \label{spin}
\end{figure}
The magnetic moments of the structures can be understood as a
consequence of graphene being a bipartite lattice in the nearest
neighbor approximation as shown by Lieb \cite{Lieb1989}. According
to Lieb's theorem \cite{Lieb1989}, the total magnetic moment can be
written as $M=N_A-N_B$ where $N_{A(B)}$ is the number of atoms
occupying the $A(B)$ sites of the bipartite graphene lattice. Thus,
if the angle between the zigzag edges is $0^\circ$ or $60^\circ$ the
edge-atoms belong to the same sublattice, while they belong to
different sublattices if the angle is $120^\circ$ or $180^\circ$.
Consequently, the hexagonal hole is non-magnetic and the triangular is
magnetic. This is consistent with a Mulliken analysis from the DFT
calculations which shows a non-zero magnetic moment only for the
$\{6,5\}_\triangle$ system of 3.00 $\mu_{B}$ per unit cell. By inspection of
the geometry we indeed find $N_A-N_B$=3.

As a continuation of Lieb's work, Inui {\it et al.} \cite{Inui1994}
showed that such sublattice imbalance results in there being
$N_A-N_B$ midgap states with zero energy. We thus expect a
degeneracy of 3 of the low-dispersion bands in the $\{6,5\}_\triangle$ case.

In Fig. \ref{bands}, lower right panel, there is one largely
dispersionless band  just below two almost completely degenerate
bands which have some dispersion especially at the $\Gamma$-point.
Such band curvature was also found for hydrogenated graphene ribbons
by Kusakabe {\it et al.} \cite{Kusakabe2003} The $\Gamma$-point
states for each band are shown in Fig. \ref{prob}, where the strong
localization is  seen for the lowest band, Fig. \ref{prob}(c),
whereas the bands with curvature, Fig. \ref{prob}(a,b), yield less
localized states. Note also the alternation in the amplitudes of the
states between sublattices as proposed by Inui {\it et al.}
\cite{Inui1994} The state of the highest occupied (spin-degenerate)
band for the non-magnetic $\{6,3.6\}_\bigcirc$ structure is shown for
comparison in Fig. \ref{prob}(d). The splitting as well as the
curvature is less pronounced for the unfilled states above the Fermi
level showing particle-hole asymmetry. This asymmetry is expected
due to the breaking of the symmetry of the bipartite lattice partly
due to the DFT treatment beyond nearest neighbor as well as the
passivation of the edges which changes the on-site potential at
edge-sites. This is inherently also the case for DFTB. By inspection
of the SIESTA Hamiltonian\cite{siesta} we find an increase in
on-site energy for passivated edge atoms as compared to atoms far
from the edge. Vanevic {\it et al.} \cite{arxiv.0903.0918v1} find
that a potential shift on the edge-atoms mainly causes a lifting of
the degeneracy of the flat bands, consistent with our
observations.

As mentioned above, the global spin is given by the sublattice
imbalance. This does not, however, determine the local spin. For the
hexagonally shaped hole structures we find not only a zero global
spin, but also a zero spin on all atoms. This explains the identical
band gaps found using DFT with and without spin. Such non-magnetic
solutions have been found for finite graphene ribbons or graphene
flakes by Jiang {\it et al.} \cite{Jiang2007} using DFT. They find a
sudden transition from non-magnetic to magnetic solutions going from
a system size of [3,3] and [4,3] with numbers indicating rings in
the graphene lattice along zig-zag and armchair directions. Such
transitions are also seen in slits cut in graphene in a study by
Kumazaki {\it et al. }\cite{Kumazaki2008} Viewing each edge in our hexagonal structures as ribbons, the largest ribbon
corresponds to a [4,3] ribbon ($\{6,3.6\}_\bigcirc$ structure). We thus expect local
magnetization to appear for slightly larger systems.

For $\{6,5\}_\triangle$ we have the strongest polarization at the middle of
each edge with maximum magnetic moment per atom being 0.24 $\mu_B$.
We note that each corner atom has a magnetization -0.03 $\mu_B$. The edge atom
magnetization is below the maximum of 1/3 $\mu_B$ for graphene
ribbons when the width becomes too large for edge-edge interactions.
Our systems thus have edge-edge interactions which is expected due
to the ribbon width of 6 rings. A plot of the spin-polarized
density\footnote{Due to the larger cell used in this calculation we
use less accurate settings (single-$\zeta$ basis, 100 Ry mesh, one $k$-point).
This changes the magnitude of the spin somewhat as compared to the
smaller cell calculations. The maximum magnetic moment per atom is
0.27 $\mu_B$ for the large cell versus 0.24 $\mu_B$ for the smaller.
The qualitative picture of the spin distribution is, however, the
same for both calculations.} is shown in Fig. \ref{spin}. The
majority spins reside mostly on the edges of the dominant sublattice
sites. Neighboring sites on the other sublattice have minority
spin-polarization. Note also the non-zero spin of the atoms far from
the edges indicating interaction between neighboring hole-edges.
\section{Conclusion}\label{sec_conc}
Using DFT and DFTB we have calculated band gaps in various antidot
lattice geometries.  The computed band gaps range from 0.2 to 1.5
eV. In general, DFTB gives larger band gaps than DFT with the largest
difference for non-magnetic structures of 44 \%. Geometry relaxation
using either method is found to increase the band gap with maximally
15\%. Combining the two methods by performing a DFT-calculation on a
DFTB-relaxed structure is found to give a good trade-off between
accuracy and computational cost facilitating the treatment of larger
systems. However, even for unrelaxed geometries we find qualitative
agreement with the DFT-relaxed geometries. Trends for ideal
geometries as presented here can thus be investigated without any
relaxation in the non-magnetic case.

Certain geometries are shown with DFT to have a non-zero total
magnetic moment which is understood via Lieb's theorem as a
consequence of sublattice imbalance. For these structures a
spin-polarized treatment is needed to achieve even qualitative
results for the band gaps. Local spin, which can occur regardless of
the total magnetic moment, is not observed for very edge-dominated
systems. Sublattice imbalance leads to the occurrence of
low-dispersion midgap bands. We find a lifting of degeneracy of
these otherwise degenerate bands  on a perfect bipartite lattice as
well as a considerable spin-splitting. These completely
spin-polarized states are primarily located at the hole-edges.

\begin{acknowledgments}
We thank Jesper G. Pedersen for fruitful discussions. Financial
support from Danish Research Council FTP grant 'Nanoengineered
graphene devices' is gratefully acknowledged. Computational
resources were provided by the Danish Center for Scientific
Computations (DCSC). APJ is grateful to the FiDiPro program of the
Finnish Academy.
\end{acknowledgments}

\bibliography{./final}

\end{document}